\documentclass[12pt,a4wide,3p,authoryear,nopreprintline]{article}
\usepackage{natbib}
\usepackage[]{geometry}
\usepackage{a4wide}
\usepackage{enumerate}
\usepackage{float}
\usepackage{caption}

\usepackage{parskip}

\usepackage{color}
\usepackage{graphicx}
\usepackage{booktabs}
\usepackage{capt-of}
\usepackage{floatpag}
\usepackage[utf8x]{inputenc}  
\usepackage{amsmath,amsfonts,amssymb,amsthm}
\usepackage{mathtools} 
\usepackage{dsfont}
\usepackage{rotating}
\usepackage{multicol}

\usepackage{url}

\usepackage[font=small,labelfont=bf]{caption}

\DeclareMathOperator*{\argmin}{arg\,min}

\usepackage{hyperref}

\allowdisplaybreaks

\begin{document} 
\vspace*{3em}
\begin{center}
\begin{LARGE}
How to Guide Decisions with Bayes Factors
\end{LARGE}\\[1.5em]
\begin{large}
Patrick Schwaferts, Thomas Augustin
\end{large}\\[1em]
\begin{scriptsize}
patrick.schwaferts@stat.uni-muenchen.de\\
thomas.augustin@stat.uni-muenchen.de\\[1em]
Ludwig-Maximilians-Universität Munich\\
Department of Statistics\\
Methodological Foundations of Statistics and its Applications\\
Ludwigsstraße 33, 80539 Munich, Germany\\[5em]
\end{scriptsize}
\end{center}

\begin{abstract}
Some scientific research questions ask to guide decisions and others do not. By their nature frequentist hypothesis-tests yield a dichotomous test decision as result, rendering them rather inappropriate for latter types of research questions. Bayes factors, however, are argued to be both able to refrain from making decisions and to be employed in guiding decisions. This paper elaborates on how to use a Bayes factor for guiding a decision. In this regard, its embedding within the framework of Bayesian decision theory is delineated, in which a (hypothesis-based) loss function needs to be specified. Typically, such a specification is difficult for an applied scientist as relevant information might be scarce, vague, partial, and ambiguous. To tackle this issue, a robust, interval-valued specification of this loss function shall be allowed, such that the essential but partial information can be included into the analysis as is. Further, the restriction of the prior distributions to be proper distributions (which is necessary to calculate Bayes factors) can be alleviated if a decision is of interest. Both the resulting framework of hypothesis-based Bayesian decision theory with robust loss function and how to derive optimal decisions from already existing Bayes factors are depicted by user-friendly and straightforward step-by-step guides.\\

\end{abstract}

\section{Introduction}

The result of a classic frequentist hypothesis test is a dichotomous test decision. However, scientific research questions are very versatile and there is not always the demand to guide a decision. By the their nature, frequentist hypothesis tests prohibit a statistical hypothesis-based analysis without making decisions. In that sense, a statistical framework that provides results without requiring an underlying (potentially artificially constructed) decision problem seems to be advantageous.
The Bayes factor -- a Bayesian quantity that is used for hypothesis comparisons \citep{Jeffreys1961,Kass1995,Goenen2005,Rouder2009} -- is argued to do so, as it is typically interpreted as evidence quantification w.r.t.\ the contrasted hypotheses \citep[see e.g.][]{Morey2016} without requiring a decision to be made. In this regard, \citet{Rouder2018} state that ``[r]efraining from making decisions strikes [them] as advantageous in most contexts.''

Naturally, the evidence (as quantified by the Bayes factor) might then be used to update beliefs in the considered hypotheses and subsequently to guide a respective decision. The essential point, however, is that the researcher might stop the analysis after calculating a Bayes factor without guiding a decision, e.g.\ if merely the evidence quantification is of interest. Then the result of the Bayesian analysis is the Bayes factor itself and not a decision.
Yet, for those research situations that do indeed aim at guiding a decision, the Bayes factor might naturally be used to do so. The aim of this elaboration is to outline the decision theoretic framework in which Bayes factors are involved.

Further, it shall be acknowledged that the specification of the relevant quantities within such a decision theoretic framework as precise values might not always be possible for an applied scientist, as the available relevant information might be scarce, vague, partial, and ambiguous. To tackle this issue, also a robust version of the framework shall be outlined in which the applied researcher is allowed to specify the essential quantities less precisely as sets of values, such that the partial nature of the relevant information might be captured more accurately. Although such robust specifications might be possible for all essential quantities \citep[see e.g.][]{Schwaferts2019,Schwaferts2021}, the present elaboration is restricted to a robustly specified inverval-valued loss function, as it is this quantity which characterizes the difference between a decision-theoretic and a non-decision-theoretic analysis, yet its precise specification is expected to bear serious difficulties for applied scientists.

The elaborations within this paper are structured as follows: After delineating the general (Section~\ref{sec:decision}) and the hypothesis-based (Section~\ref{sec:hypo_decision}) framework of Bayesian decision theory, its relation with Bayes factors is depicted (Section~\ref{sec:BF}). To facilitate a more user-friendly employment of the hypothesis-based Bayesian decision theoretic framework, a robust interval-valued specification of the loss function was allowed (Section \ref{sec:robust_loss}) and the restriction of the prior distributions to be proper can be alleviated (Section~\ref{sec:improper_priors}). Both the resulting framework (Secion~\ref{sec:guide1}) and how to derive optimal actions from existing Bayes factor values (Secion~\ref{sec:guide2}) are presented in respective step-by-step guides.

\section{Bayesian Decision Theory}\label{sec:decision}

Within the framework of Bayesian decision theory \citep[e.g.][]{Berger1985,Robert2007}, the objective is to decide between different actions. In accordance with the context of Bayes factors, only two actions shall be considered, namely $a_0$ and $a_1$, being comprised within the action space $\mathcal{A}= \{ a_0 , a_1 \}$.

The researcher plans to conduct an investigation that yields data $\boldsymbol{x}$, which is characterized by a parametric sampling distribution with parameter $\theta \in \Theta$, where $\Theta$ is the parameter space. Accordingly, the density of the data is $f(\boldsymbol{x}|\theta)$.

In a Bayesian setting, a prior distribution on the parameter $\theta$ with density $\pi(\theta)$ needs to be specified. This prior reflects the information (or belief or knowledge or uncertainty) about the parameter before the investigation is conducted.

In addition, also a loss function $L: \Theta \times \mathcal{A} \to \mathbb{R}_0^+ : (\theta,a) \mapsto L(\theta,a)$ needs to be specified, which quantifies the ``badness'' of the consequences of deciding for the action $a \in \mathcal{A}$ if the parameter value $\theta \in \Theta$ is true. Usually, the exact shape of this loss function is inaccessible and hypothesis-based analyses are able to tackle this issue. These are depicted within the next section, but first -- to delineate the ideal Bayesian solution -- assume $L$ is fully known.

Now, after specifying the parametric sampling distribution, the prior, as well as the loss function, the investigation can be conducted and the data $\boldsymbol{x}$ are observed.  
This allows to update the prior distribution via Bayes rule to the posterior distribution with density
\begin{equation}
\pi(\theta|\boldsymbol{x}) = \frac{f(\boldsymbol{x}|\theta) \, \pi(\theta)}{f(\boldsymbol{x})} = \frac{f(\boldsymbol{x}|\theta) \, \pi(\theta)}{\int_\Theta f(\boldsymbol{x}|\theta) \, \pi(\theta) \, d\theta}   \, .
\end{equation}
There are plenty of resources available about how to obtain this posterior \citep[e.g.][]{Gelman2013,Kruschke2015}, which reflects the information (or belief or knowledge or uncertainty) about the parameter after the investigation was conducted.

Based on this posterior distribution, it is possible to calculate the expected posterior loss $\rho:\mathcal{A} \to \mathbb{R}_0^+$ for each action by integrating the loss function $L$ over the posterior density:
\begin{equation}
\rho(a) = \int_\Theta L(\theta,a) \,  \pi(\theta|\boldsymbol{x}) \, d\theta \, .
\end{equation}
The optimal action $a^*$ has minimal expected posterior loss:
\begin{equation}
a^* = \argmin_{a \in \mathcal{A}} \rho(a) \, .
\end{equation}

\section{Hypothesis-Based Bayesian Decision Theory}\label{sec:hypo_decision}

As mentioned, typically, the loss function $L$ is not fully accessible as the essential information about it might be scarce, vague, partial, and ambiguous. A commonly employed solution is a hypothesis-based analysis: The researcher considers each possible parameter value $\theta$ and assesses which action should be preferred if this parameter value would be true.
These considerations lead to two sets of parameters $\Theta_0$ and $\Theta_1$ for which the actions $a_0$ and $a_1$ should be preferred, respectively. These sets define the hypotheses
\begin{equation} \label{eq:hypo}
H_0: \theta \in \Theta_0 \quad \text{vs.} \quad H_1: \theta \in \Theta_1
\end{equation}
employed in conventional analyses, such as hypothesis tests or Bayes factors.

From the posterior density $\pi(\theta|\boldsymbol{x})$ it is possible to determine the posterior probabilities of the parameters sets $\Theta_0$ and $\Theta_1$, i.e.\ of the hypotheses $H_0$ and $H_1$, by
\begin{equation} \label{eq:post_belief_hypo}
P(H_0|\boldsymbol{x}) = \int_{\Theta_0} \pi(\theta|\boldsymbol{x}) \, d\theta
\quad \quad \text{and} \quad \quad
P(H_1|\boldsymbol{x}) = \int_{\Theta_1} \pi(\theta|\boldsymbol{x}) \, d\theta \, ,
\end{equation}
respectively.
The ratio of these beliefs $P(H_0|\boldsymbol{x}) /P(H_1|\boldsymbol{x})$ is referred to as posterior odds.

\begin{figure}[ht]
	\centering
  \includegraphics[width=1\textwidth]{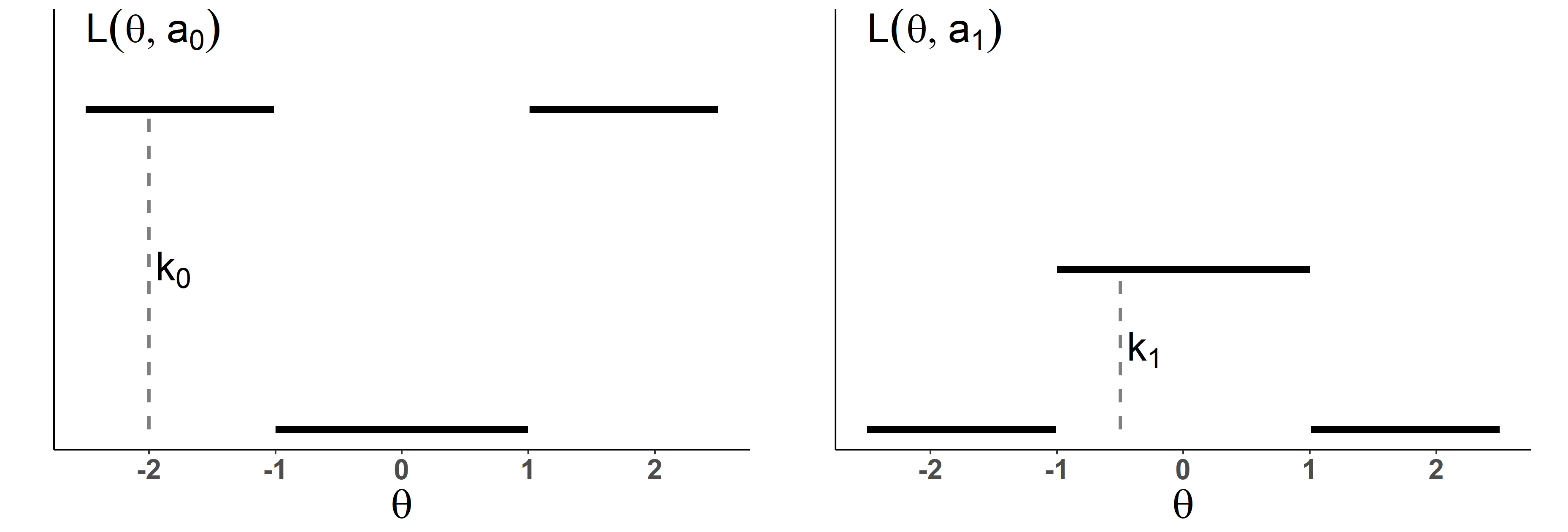}
	\caption{Hypothesis-Based Loss Function. Assume $\Theta = \mathbb{R}$, $\Theta_0 = [-1,1]$, and $\Theta_1 = (-\infty,-1) \cup (1,\infty)$. The hypothesis-based loss function $L$ (y-axis) in regret form (see Table~\ref{tab:loss}) in dependence of the parameter $\theta$ (x-axis) and the actions $a_0$ (left) and $a_1$ (right) is assumed to be constant within the sets $\Theta_0$ and $\Theta_1$, respectively. This is an assumption (\textit{simplification assumption}) inherent to a hypothesis-based statistical analysis which -- at least implicitly -- considers an underlying applied decision problem.}
	\label{fig:loss}
\end{figure}

The underlying assumption of hypothesis-based analyses is that the loss values within these sets $\Theta_0$ and $\Theta_1$ are constant, respectively (see Figure~\ref{fig:loss}). This assumption shall be referred to as \textit{simplification assumption} and is inherent to a statistical analysis which considers statistical hypotheses and derives applied conclusions based on respective (hypothesis-based) results.
In addition (without loss of generality), the loss values for deciding correctly (i.e.\ for $a_0$ if $\theta \in \Theta_0$ or for $a_1$ if $\theta \in \Theta_1$) can be set to $0$. The resulting loss function is in regret form (depicted in Table~\ref{tab:loss}) and has only two values to specify: $k_0 := L(a_0, \theta)$ if $\theta \in \Theta_1$ and $k_1 := L(a_1, \theta)$ if $\theta \in \Theta_0$.

\begin{table}[]
\centering
\caption{Simplified Hypothesis-Based Loss Function.}
	\label{tab:loss}
\begin{tabular}{c|cc}
$L(\theta,a)$ & $\theta \in \Theta_0$ & $\theta \in \Theta_1$ \\ \hline
$a = a_0$ & $0$ & $k_0$ \\
$a = a_1$ & $k_1$ & $0$
\end{tabular}
\end{table}

With this simplified loss function (Table \ref{tab:loss}), the expected posterior loss of each action can be calculated as
\begin{align}
\rho(a_0) &= \int_\Theta L(\theta,a_0) \,  \pi(\theta|\boldsymbol{x}) \, d\theta = k_0 \cdot P(H_1|\boldsymbol{x}) \\
\rho(a_1) &= \int_\Theta L(\theta,a_1) \,  \pi(\theta|\boldsymbol{x}) \, d\theta = k_1 \cdot P(H_0|\boldsymbol{x}) \, 
\end{align}
and the action with minimal expected posterior loss shall be selected.

Only the ratio $k := k_1/k_0$ is required to determine this optimal action. This ratio $k$ states how much worse it would be to decide for $a_1$ if $\theta \in \Theta_0$ is true (type-I-error) than to decide for $a_0$ if $\theta \in \Theta_1$ is true (type-II-error), if deciding correctly has loss $0$. With the ratio of expected posterior losses
\begin{equation}
\varrho(k) := \frac{\rho(a_1)}{\rho(a_0)} = k \cdot \frac{P(H_0|\boldsymbol{x})}{P(H_1|\boldsymbol{x})}
\end{equation}
the optimal action is
\begin{equation}
a^* =
\begin{cases}
a_0 \quad \text{if} \quad \varrho(k) > 1 \\
a_1 \quad \text{if} \quad \varrho(k) < 1 
\end{cases} \, .
\end{equation}

For $\varrho(k) = 1$ any action might be chosen.

\section{Bayes Factors}\label{sec:BF}

\begin{figure}[ht]
	\centering
  \includegraphics[width=1\textwidth]{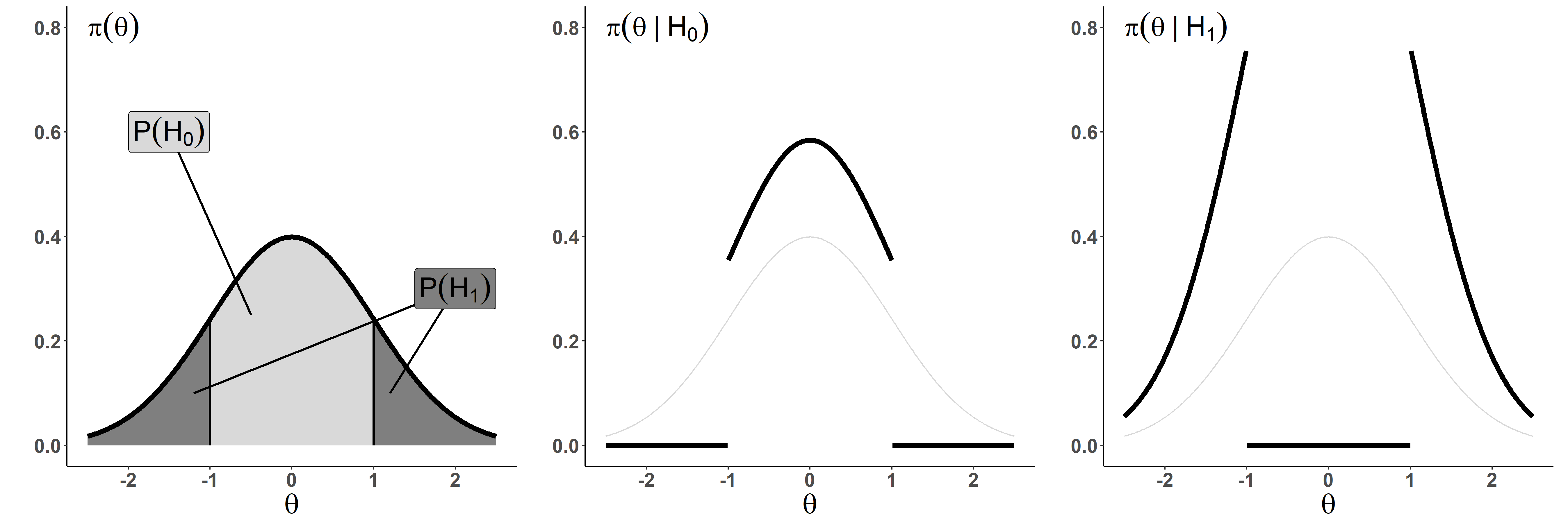}
	\caption{Prior Decomposition. Assume $\Theta = \mathbb{R}$, $\Theta_0 = [-1,1]$, $\Theta_1 = (-\infty,-1) \cup (1,\infty)$ and a standard normal distribution for $\theta \sim N(0,1)$. Left:	 The prior density $\pi(\theta)$ is depicted as solid line. $P(H_0)$ and $P(H_1)$ can be calculated as respective areas under this density, depicted as light gray and dark gray, respectively. Middle: The within-hypothesis density $\pi(\theta|H_0)$ as in equation~(\ref{eq:hypo_prior_0}) is depicted as solid line. Right: The within-hypothesis density $\pi(\theta|H_1)$ as in equation~(\ref{eq:hypo_prior_1}) is depicted as solid line. }
	\label{fig:prior_decomposition}
\end{figure}

By assessing even the prior distribution in the light of the hypotheses, it is possible to obtain the prior probabilities in the hypotheses (illustrated in Figure~\ref{fig:prior_decomposition}, left): 
\begin{equation}
P(H_0) = \int_{\Theta_0} \pi(\theta) \, d\theta
\quad \quad \text{and} \quad \quad
P(H_1) = \int_{\Theta_1} \pi(\theta) \, d\theta \, ,
\end{equation}
Analogously, the ratio of these beliefs $P(H_0) /P(H_1)$ is referred to as prior odds.

In addition, the prior distribution can be restricted to each of the hypotheses, referred to as within-hypothesis priors (illustrated in Figure~\ref{fig:prior_decomposition}, middle and right), and the corresponding densities are
\begin{align}
\pi(\theta|H_0) &= \frac{1}{P(H_0)} \, \pi(\theta) \cdot \mathsf{1}(\theta \in \Theta_0) \label{eq:hypo_prior_0}\\
\pi(\theta|H_1) &= \frac{1}{P(H_1)} \, \pi(\theta) \cdot \mathsf{1}(\theta \in \Theta_1) \, ,\label{eq:hypo_prior_1}
\end{align}
where $\mathsf{1}(s) = 1$ if the statement $s$ is true and $\mathsf{1}(s) = 0$ if $s$ is false.

The overall prior distribution can be decomposed \citep[cp.][]{Rouder2018} into the prior probabilities of the hypotheses and the within-hypothesis priors (Figure~\ref{fig:prior_decomposition}):
\begin{equation}
\pi(\theta) = P(H_0) \, \pi(\theta|H_0) + P(H_1) \, \pi(\theta|H_1) \, .
\end{equation}

Instead of considering the overall prior distribution together with the hypotheses (which leads to the posterior odds, as in Section~\ref{sec:hypo_decision}), the Bayes factor is obtained by considering only the within-hypothesis priors together with the hypotheses:
\begin{equation} \label{eq:BF}
B\!F := \frac{\int_{\Theta_0} f(\boldsymbol{x}|\theta) \, \pi(\theta|H_0) \, d\theta}{\int_{\Theta_1} f(\boldsymbol{x}|\theta) \, \pi(\theta|H_1) \, d\theta} \, .
\end{equation}

The posterior odds can then be calculated from the Bayes factor and the prior odds:
\begin{equation}\label{eq:post_odds}
\frac{P(H_0|\boldsymbol{x})}{P(H_1|\boldsymbol{x})} = B\!F \cdot \frac{P(H_0)}{P(H_1)} \, .
\end{equation}
The optimal decision can now be obtained as in the previous section (Section~\ref{sec:hypo_decision}) by considering the loss function.

\section{Robust Loss Function}\label{sec:robust_loss}

However, a precise specification of the value $k$ is typically not accessible, as essential information about the ``badness'' of the consequences of the decision are scarce, vague, partial, and ambiguous.
Yet, this partial information needs to be included into the analysis, as ignoring it facilitates suboptimal decisions. A decision cannot be guided properly without considering its consequences.

This partial information about the loss can be captured less arbitrarily and more robustly by an interval $[\underline{K},\overline{K}]$ than by a precise value $k$ \citep[cp.\ e.g.][]{Walley1991, Augustin2014}. To do so, the researcher has to determine a lower bound $\underline{K}$ and an upper bound $\overline{K}$ for reasonable $k$ values (i.e.\ for the ratio of how much worse the type-I-error is compared to the type-II-error, if deciding correctly has a loss of $0$).

To perform a robust analysis \citep[cp.\ also][]{RiosInsua2012} with this interval-valued specification, it is possible to obtain and consider the optimal action for each value within this interval $[\underline{K},\overline{K}]$.

If the optimal action is the same for each $k$ within $[\underline{K},\overline{K}]$, then this action should be chosen.
If not, the decision should be withheld, because the data or the information about the decision problem are not sufficient to unambiguously guide the decision.

Formally \citep[see also][]{Schwaferts2019, Schwaferts2020, Schwaferts2021}, the ratios of expected posterior losses need to be calculated for both the lower and upper bound, respectively:
\begin{equation} \label{eq:ratio_imprecise}
\varrho(\underline{K}) = \underline{K} \cdot \frac{P(H_0|\boldsymbol{x})}{P(H_1|\boldsymbol{x})}
\quad \quad \text{and} \quad \quad
\varrho(\overline{K}) = \overline{K} \cdot \frac{P(H_0|\boldsymbol{x})}{P(H_1|\boldsymbol{x})} \, .
\end{equation}

Then, the optimal action is
\begin{equation} \label{eq:optimal_decision_imp}
a^* =
\begin{cases}
a_0 \quad \text{if} \quad \varrho(\underline{K}) \geq 1 \\
a_1 \quad \text{if} \quad \varrho(\overline{K}) \leq 1 
\end{cases} \, .
\end{equation}

For $\varrho(\underline{K}) < 1 < \varrho(\overline{K})$, the decision should be withheld.

\section{Improper Priors}\label{sec:improper_priors}

\begin{figure}[ht]
	\centering
  \includegraphics[width=0.9\textwidth]{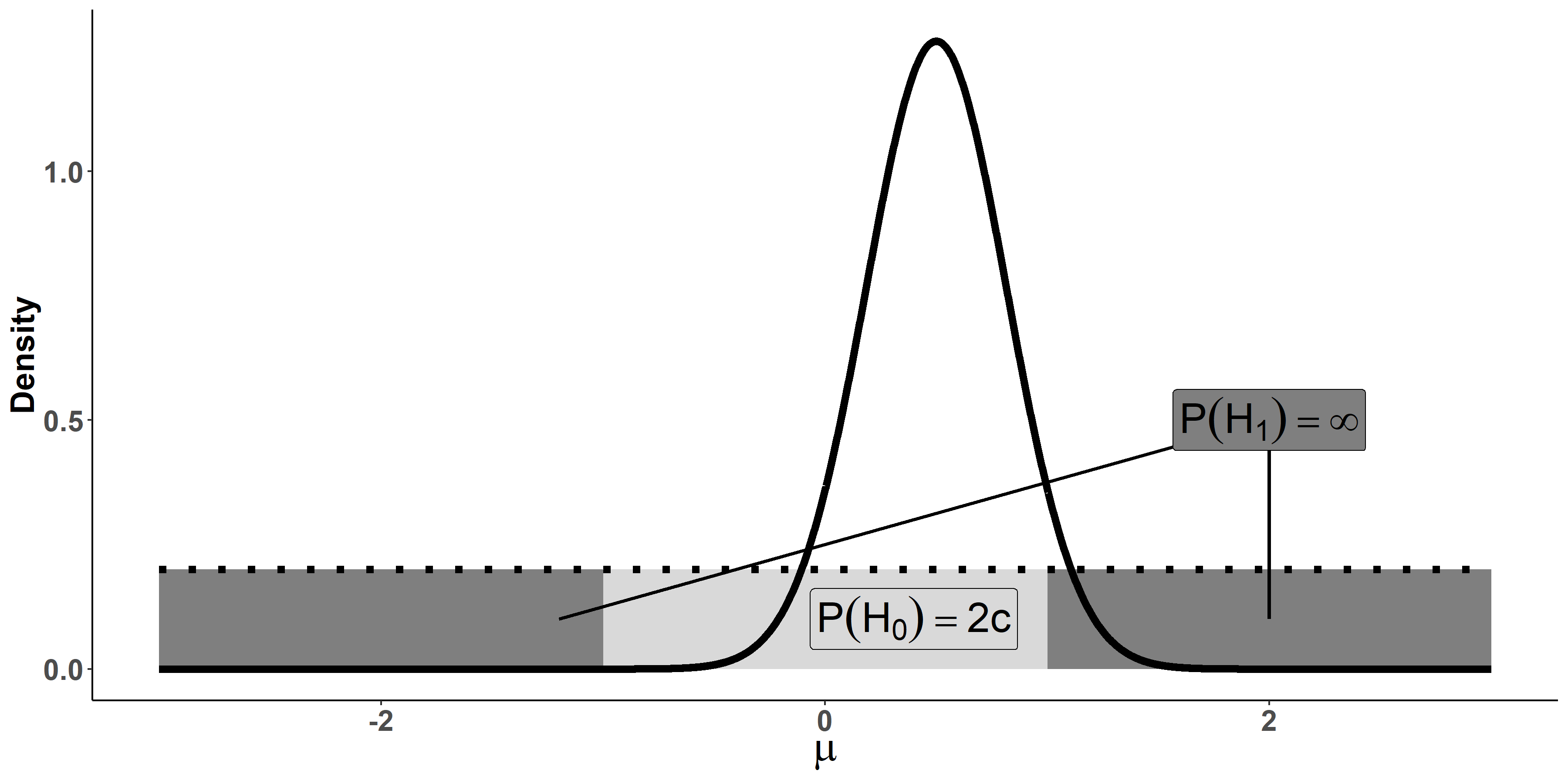}
	\caption{Improper Prior. Assume the model $X_i \overset{iid}{\sim} N(\mu,\sigma^2)$ for $i=1,\dots,n$, with known variance $\sigma^2=1$ and unknown parameter $\mu \in \mathbb{R}$, the hypotheses $\Theta_0 = [-1,1]$, $\Theta_1 = (-\infty,-1) \cup (1,\infty)$. The function $\pi(\mu) = c$, with $c=0.2$ being an arbitrary constant, characterizes the improper prior distribution for $\mu$ (dotted line). For a sample of size $n=10$ with in-sample mean $\bar{x} = 0.5$, the posterior distribution is proper (solid line), such that its density integrates to $1$. The prior ``beliefs'' into the hypotheses are with $P(H_0) = 2c$ and $P(H_1) = \infty$ not reasonably interpretable (light gray and dark gray, respectively).}
	\label{fig:improper_prior}
\end{figure}

Furthermore, the calculation of Bayes factors comes along with a restriction \citep[][]{Jeffreys1961} on the prior distribution: It must be a proper distribution, i.e.\ the density has to integrate to $1$:
\begin{equation} \label{eq:proper_density}
\int_\Theta \pi(\theta) \, d \theta = 1 \, .
\end{equation}
In contrast, an improper prior distribution is characterized by a non-integrable function (e.g.\ $\pi(\theta) \propto c$ with $c >0$ being a constant, see Figure~\ref{fig:improper_prior}, dotted line)  and, technically, this prior distribution is no proper probability distribution. However, these improper priors are frequently allowed within Bayesian prior specifications, because they might lead to proper posterior distributions (see Figure~\ref{fig:improper_prior}, solid line). In this case, the posterior odds $P(H_0|\boldsymbol{x})/P(H_1|\boldsymbol{x})$ can be calculated reasonably and a decision can be guided consistently.

The prior odds, however, might not be reasonable (e.g.\ with $P(H_0)/P(H_1) = 0$ as in Figure~\ref{fig:improper_prior}). Accordingly, the Bayes factor (calculated via equation~(\ref{eq:post_odds}))
\begin{equation}
B\!F = \left. \frac{P(H_0|\boldsymbol{x})}{P(H_1|\boldsymbol{x})} \middle/ \frac{P(H_0)}{P(H_1)} \right.
\end{equation}
cannot be calculated reasonably due to its dependence on the prior odds. Therefore, Bayes factors require -- in contrast to a Bayesian analysis in general -- proper prior distributions. This is truly a limitation, as improper priors are frequently employed for representing non-knowledge or for letting the data speak for themselves \citep[e.g.][]{Gelman2013}.

This issue is alleviated in hypothesis-based Bayesian decision theoretic accounts, as improper priors typically yield proper posterior odds. Accordingly, a researcher who is interested in guiding a decision might employ the decision theoretic framework directly without explicitly calculating the Bayes factor. Then, also improper priors might be employed.

Please note that it is still an ongoing debate whether non-knowledge can be formalized by a precise improper prior distribution and if so, which improper prior distribution shall be employed. Although the authors of this paper doubt that non-knowledge can be formalized by a precise prior distribution, even if it is improper \citep[cp.\ e.g.][]{Augustin2014}, this issue shall not be addressed here. In general, it is important that the employed prior distribution matches with the available information (or non-information) about the phenomenon of interest, and this applies to every point of view within this debate. In this regard, the present elaboration emphasizes only that it is mathematically possible to employ improper priors if decisions are of interest, which is an advantage of the (more general) hypothesis-based Bayesian decision theoretic account over Bayes factors.

\section{Step-By-Step Guides}\label{sec:guides}

\subsection{Hypothesis-Based Bayesian Decision Theory}\label{sec:guide1}

In order to apply this hypothesis-based Bayesian decision theoretic framework with robust loss function, a researcher might follow the following steps.

\textbf{Step 1: Actions.} First of all, the researcher needs to specify the actions. It is recommended to explicitly state and report these actions, e.g.\ by \citep[this example is taken from][]{Bartolucci2011}
\begin{itemize}
\item[$a_0$:] do not administer aspirin to prevent myocardial infarction
\item[$a_1$:] administer aspirin to prevent myocardial infarction
\end{itemize}
If the researcher has difficulties stating the actions, maybe there is no decision to guide and a descriptive analysis might suffice \citep[cp.\ also][]{Cumming2014,Kruschke2018b}.

\textbf{Step 2: Sampling Distribution.} Next, the researcher should provide a detailed description of the investigation and how it is characterized (i.e.\ the sampling distribution). It is recommended to also explicitly state the employed parameter $\theta$ and its interpretation. This is the basis for specifying the hypotheses.

\textbf{Step 3: Prior Distribution.} In the Bayesian setting, it is possible to include prior information (or belief or knowledge or uncertainty) into the analysis. In that, the researcher has to specify a prior distribution on the parameter. It is recommended to fully report the available prior information about the parameter $\theta$ and why this leads to the prior density $\pi(\theta)$.

Of course, this specification is far from being unambiguous. However, this is a fundamental issue inherent to every Bayesian analysis (not only Bayesian decision theoretic accounts) and solving this issue is not the intention of this elaboration. Nevertheless, solutions, such as sensitivity analyses \citep[found in almost every Bayesian textbook, e.g.][]{Gelman2013}, exist. It is recommended at this step of the analysis to also state all other possible prior densities that are in accordance with the available prior information, as these serve as basis for a subsequent sensitivity analysis.

Naturally, also non-informative priors might be specified and they might also be improper (as long as they lead to proper posterior distributions).

\textbf{Step 4: Assumption.} If the researcher is unable to specify the loss function $L$, then a hypothesis-based simplification as in Section \ref{sec:hypo_decision} might be a solution. This simplification is an assumption on the loss function, namely that the loss function is constant within each of two parameter sets.
If this assumption is not appropriate, it might lead to errors (which are inherent to every hypothesis-based analysis) and the researcher needs to be aware of this consequence. It is recommended to explicitly report that this assumption was made. Transparency is one of the basic principles in science \citep[cp.][]{Gelman2017}.	

\textbf{Step 5: Hypotheses.}  Now, the researcher has to consider each possible parameter value $\theta$ and assess which action should be preferred if this parameter value would be true.
All parameters for which $a_0$ or $a_1$ should be preferred are comprised within the sets $\Theta_0$ or $\Theta_1$, respectively.
Certainly, there are parameter values that define the border between both sets $\Theta_0$ and $\Theta_1$. It is recommended to explicitly state what these values mean in real-life and why they define reasonable borders between $\Theta_0$ and $\Theta_1$.

\textbf{Step 6: Errors.}
Deciding for $a_1$ if $\theta \in \Theta_0$ is the type-I-error and deciding for $a_0$ if $\theta \in \Theta_1$ is the type-II-error. Both errors should be delineated, as they serve as basis for specifying the ratio $k$. It is recommended to explicitly state these errors and their consequences, e.g.\ by
\begin{itemize}
\item[] Type-I-error: administer aspirin to prevent myocardial infarction, but the effect is negligible. Consequence: patients unnecessarily suffer side effects of aspirin.
\item[] Type-II-error: do not administer aspirin to prevent myocardial infarction, although it would have an effect. Consequence: some patients suffer a myocardial infarction, which could have been prevented.
\end{itemize}
Of course, this is only a schematic illustration and in real empirical studies these elaborations will be more comprehensive.

\textbf{Step 7: Loss Magnitude.} The researcher has to imagine that the ``badness''  of deciding correctly is $0$. In this context, the researcher has to determine how much worse the type-I-error is compared to the type-II-error. This is the value $k$. As a precise value for $k$ is difficult to determine, it might be easier to specify a range $[\underline{K},\overline{K}]$ of plausible values for $k$. It is recommended to report all considerations that lead to this specification.

\textbf{Step 8: Investigation.} Now, the investigation can be conducted and it is recommended to preregister\footnote{Study designs can be preregistered e.g.\ at \url{www.cos.io/initiatives/prereg}.} the previous specifications, the design of the experiment, and the planned (decision theoretic) analysis of the data \citep[cp.][]{Nosek2018,Klein2018}. Registered reports\footnote{Information about registered reports can be found e.g.\ at \url{www.cos.io/rr}.} even allow to obtain a peer-review prior to collecting the data.

\textbf{Step 9: Posterior Distribution.} The observed data are used to obtain the posterior distribution as well as the posterior beliefs in the hypotheses $P(H_0|\boldsymbol{x})$ and $P(H_1|\boldsymbol{x})$. There are countless references on how to do this \citep[e.g.][]{Gelman2013,Kruschke2015}.

\textbf{Step 10: Optimal Action.} The researcher has to calculate $\varrho(\underline{K})$ and $\varrho(\overline{K})$ as in equation~(\ref{eq:ratio_imprecise}) to find the optimal action as in equation ~\ref{eq:optimal_decision_imp}).

For $\varrho(\underline{K}) < 1 < \varrho(\overline{K})$, the decision should be withheld, because the data or the information about the decision problem are not sufficient to unambiguously guide the decision. In this case, a reasonable strategy might be to collect more data or to gather more information about the decision problem, especially about the consequences of the errors, to narrow down $[\underline{K},\overline{K}]$. However, it is recommended to transparently report that a decision was withheld at first and which subsequent steps were taken to obtain more information.

\textbf{Step 11: Publish Data.} Of course, other researchers might need the data to guide their decisions. It is to expect that they have different prior knowledge and that their decisions employ different hypotheses. Without having access to the data set (but only to the reported analysis), it might be difficult, or even impossible, for them to guide their decisions properly, emphasizing the importance of open science\footnote{Comprehensive information about open science are provided e.g.\ by the LMU Open Science Center: \url{www.osc.uni-muenchen.de}.}.

\subsection{From Bayes Factors to Decisions}\label{sec:guide2}

Sometimes, a researcher wants to use the results of a previous study to guide a decision. Assume a Bayes factor $B\!F$ was already calculated and shall now be used to guide this decision.

\textbf{Step A: Applicability of the Sampling Distribution.} Confirm that the interpretation of the parameter $\theta$ is actually relevant for the decision of interest. If this is not the case, the available data (or Bayes factor) can hardly be used to guide the decision of interest.

\textbf{Step B: Applicability of the Hypotheses.} Certain specific hypotheses were assumed in order to calculate the Bayes factor. These need to match with the decision problem of interest. To assess this, the potential actions of the decision problem of interest need to be delineated as in \textit{Step~1} and the parameter sets $\Theta_0$ and $\Theta_1$ need to be obtained as in \textit{Step~5}. These sets have to be equivalent to the hypotheses that were employed in the calculation of the Bayes factor. If this is not the case, it is recommended to not use this Bayes factor value and restart the decision theoretic account within the previous section (Section~\ref{sec:guide1}). In this regard, it is helpful if the data set, that was used to calculate the original Bayes factor, is fully accessible.

\textbf{Step C: Applicability of the Prior Distribution.} Confirm that the employed within-hypothesis prior distributions for calculating the Bayes factor match with the available information about the phenomenon of interest. If this is not the case, it is recommended to not use this Bayes factor value and restart the decision theoretic account within the previous section (Section~\ref{sec:guide1}). Again, to do so it is helpful if the data set, that was used to calculate the original Bayes factor, is fully accessible.

\textbf{Step D: Prior Odds.} As the calculation of the Bayes factor does not require the prior odds, only the within-hypothesis prior distributions, former need to be specified to guide a decision. In this regard, the researcher has to specify the prior probabilities of the hypotheses. Analogue to \textit{Step~3}, as this is part of the Bayesian prior specification, it is recommended to fully report the available information about the parameter and why it leads to the prior probabilities of the hypotheses.

\textbf{Step E: Loss Function.} Specify the (interval-valued) loss function by following \textit{Steps~4, 6, and 7}. 

\textbf{Step F: Posterior Odds.} Use the available Bayes factor $B\!F$ to calculate the posterior odds via equation~(\ref{eq:post_odds}).

\textbf{Step G: Optimal Action.} The optimal action can be derived as in \textit{Step~10}.

\section{Concluding Remarks}\label{sec:conclusion}

Statisticians and methodologists do -- in general -- not know all the different fields of applications and research contexts a statistical method will eventually be employed in. The scientific endeavor is extremely versatile and research problems might arise that have not been thought of before. In that, versatility of research methods is of paramount importance. While it might be considered as disadvantageous that frequentist hypothesis tests are restricted to a dichotomous decision context, it might similarly be considered as disadvantageous if Bayes factors are restricted to only an evidential, non-decision context. Fortunately, the mathematics underlying Bayes factors suggest their involvement in guiding decisions. In this regard, Bayes factors might be seen as evidential quantification or as a quantity in the context of guiding decisions, depending on the goals of the scientific investigation.

In order to use Bayes factors correctly when guiding decision, their embedding within the framework of Bayesian decision theory has to be considered. It is important that the research context as well as the decision problem are formalized appropriately. If misspecified, results inform past the research question. Naturally, the specification of essential quantities (such as the prior distribution, the hypotheses, or the loss function) is an applied problem and might be rather difficult for the applied scientist. In order to alleviate these issues, these quantities might be specified robustly as interval-valued or set-valued quantities. Then the researcher might consider a range or a set of plausible values, avoiding the necessity to (arbitrarily) commit to one single precise value. Within this elaboration only an interval-valued loss value was considered, as it keeps the calculations simple (compare Section~\ref{sec:robust_loss}) yet allows to include essential loss information (about the consequences of the decision) into the analysis. Naturally, also the prior distribution and the hypotheses might be included into the analysis as set-valued quantities \citep[see e.g.][]{Ebner2019}. How to deal with set-valued quantities on a formal level is extensively elaborated on within the field of imprecise probabilities \citep[see e.g.][]{Walley1991, Augustin2014, Huntley2014}.

In summary, this elaboration delineates the decision theoretic embedding of Bayes factors by outlining the framework of hypothesis-based Bayesian decision theory, supplemented by considerations about robust loss specifications and straightforward step-by-step guides. These guides try to help those applied scientists who want to guide decisions with Bayes factors.

\bibliographystyle{diss-style}
\bibliography{Preprint_BFdecisions}
\end{document}